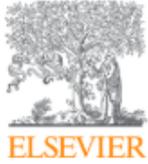
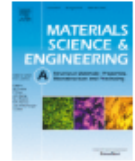

# Compressive performance and crack propagation in Al alloy/Ti$_2$AlC composites

*D.A.H. Hanaor[1] [*], L. Hu[2], W.H. Kan[1], G. Proust[1], M. Foley[3], I. Karaman[2], M. Radovic[2]*

[1] School of Civil Engineering, University of Sydney, Sydney, NSW 2006, Australia
[2] Department of Materials Science and Engineering, Texas A&M University, College Station, TX 77843, USA
[3] Australian Centre for Microscopy and Microanalysis, University of Sydney, Sydney, NSW 2006, Australia

[*] Corresponding author: dorian.hanaor@sydney.edu.au

**Abstract:**

Composite materials comprising a porous Ti$_2$AlC matrix and Al 6061 alloy were fabricated by a current-activated pressure assisted melt infiltration process. Coarse, medium and fine meso-structures were prepared with Al alloy filled pores of differing sizes. Materials were subjected to uniaxial compressive loading up to stresses of 668 MPa, leading to the failure of specimens through crack propagation in both phases. As-fabricated and post-failure specimens were analysed by X-ray microscopy and electron microscopy. Quasi-static mechanical testing results revealed that compressive strength was the highest in the fine structured composite materials. While the coarse structured specimens exhibited a compressive strength of 80% relative to this. Reconstructed micro-scale X-ray tomography data revealed different crack propagation mechanisms. Large planar shear cracks propagated throughout the fine structured materials while the coarser specimens exhibited networks of branching cracks propagating preferentially along Al alloy-Ti$_2$AlC phase interfaces and through shrinkage pores in the Al alloy phase. Results suggest that control of porosity, compensation for Al alloy shrinkage and enhancement of the Al alloy-Ti$_2$AlC phase interfaces are key considerations in the design of high performance metal/Ti$_2$AlC phase composites.

## 1. Introduction

MAX phases are a family of ternary carbides and nitrides with the formula $M_{n+1}AX_n$, where M is a transition metal, A is a group III or IV metal, X is either nitrogen or carbon and n =1, 2 or 3 [1]. This family of materials first came to attention through the work of Nowotny and his co-workers in the 1960s [2-5] and became the subject of renewed interests after 1996 with a study on synthesis and the unusual mechanical properties of bulk $Ti_3SiC_2$ [6]. Following this, the scope of this family of materials rapidly expanded, with a growing range of systems and compositions being studied. MAX phase ceramics are of great interest owing to their unusual combination of properties including high thermal and electronic conductivity, oxidation resistance, good machinability, damage tolerance, and thermal shock resilience, among others.

Among the numerous MAX phases, $Ti_2AlC$ has drawn attention and has been the subject of several investigations [7-11]. This material has been shown to exhibit beneficial properties in terms of machinability, electric conductivity and fracture toughness [9, 12-14]. Of particular interest is the crack healing ability observed in $Ti_2AlC$ systems through the formation of a well adhered alumina phase in heat treatment cycles [15-17]. Following the typical structure of materials in the $M_2AX$ subset of the MAX phase materials (also termed the 211 group), $Ti_2AlC$ comprises molecular layers of titanium carbide with every third layer consisting of pure aluminium, and belongs to the space group $P6_3/mmc$ [18]. These layers of ductile phase facilitate nonlinear kinking behaviour and plastic deformation, believed to occur through basal plane slip.

An interesting combination of metallic- and ceramic-like properties of MAX phases have motivated the development of metal/MAX phase composites [1, 19, 20]. A composite of Cu and $Ti_3SiC_2$ was proposed as a new electro-friction material [21], and MAX phases have further been shown to increase the mechanical strength of several metallic systems while maintaining good thermal and electrical conductivity [22-24]. In addition, the yield strength of a $Al/Ti_3AlC_2$ composite was found to be twice that of Al [25] while the mechanical energy dissipation was found to be significantly improved in $Mg/Ti_2AlC$ [26] and $NiTi/Ti_3SiC_2$ composites [27] when compared to its pure constituents.

Metal/MAX phase composites are frequently processed by powder co-sintering methods. One of the principal challenges in this approach stems from the reaction between the metallic and MAX phases that limits the use of high temperature processes. In powder co-sintering, temperatures are usually chosen just below the melting point of the metal phase, which is generally insufficient for the sintering of the MAX phase material [25]. In order to obtain metal/MAX phase composites with marginal inter-phase reactions, molten metals were infiltrated into MAX phase foams in a pressureless infiltration technique to fabricate $Ti_2AlC/Mg$ composites exhibiting higher strength and mechanical energy dissipation than other Mg composites [28, 29]. However, such pressureless infiltration into foams is encumbered by poor wettability of MAX phase foams by some molten metals, which inhibits adequate metal infiltration. The observed poor wettability may yield weak bonding between metal and ceramic phases, resulting in inferior mechanical properties [30]. This problem can be overcome by using pressure infiltration to force molten metals into ceramic foams. However, in many cases the reaction of the molten metal with the MAX phase material is sufficiently rapid that even in such pressure-driven infiltration methods new phases are likely to form. The new phases not only cage the pores to prevent further infiltration, but also degrade the constituents of the composites. Minimising the extent of such reactions remains a significant challenge towards the fabrication of metal/MAX phase composites.

Aluminium alloys are attractive in ceramic-metal composites for aerospace and transportation applications, where weight saving and thermal stability are important considerations. Thus Al and its alloys have been combined with ceramic phases of $Al_2O_3$ [31-33], $B_4C$ [30, 34, 35], and SiC [36, 37], in composite materials. However, MAX phases had not been used in Al-based composites until recent studies on composites of aluminium alloys with and $Ti_3AlC_2$, $Ti_2AlC$ and $V_2AlC$ [12, 38, 39] . The use of MAX phases in Al-based composites has several additional advantages relative to traditional ceramic components, *e.g.* $Al_2O_3$, $B_4C$, or SiC. Typical MAX phases exhibit a higher fracture toughness (*e.g.* ~7 $MPa·m^{1/2}$ for $Ti_3SiC_2$,) than $Al_2O_3$ (~4 $MPa·m^{1/2}$), $B_4C$ (~3.7 $MPa·m^{1/2}$), and SiC (~4.6 $MPa·m^{1/2}$). Furthermore, unlike traditional ceramics, MAX phases exhibit high thermal and electronic conductivities originating from atomic bonding with mixed covalent, ionic, and metallic characteristics, allowing for more versatile applications.

It has been shown that a current-activated, pressure-assisted infiltration (CAPAI) is a viable method for producing interpenetrating Al alloy/MAX phase composites [40]. One of the attributes of this method is that it facilitates the fabrication of composite materials that could not otherwise be fabricated using conventional methods due to poor wettability and interphase reactions [27]. Future development of composites fabricated by such means necessitates an improved understanding of multi-scale structural-mechanical property relationships in these materials. Here we present the first report of structure, mechanical performance and crack propagation in Al alloy/$Ti_2AlC$ composite systems fabricated with different meso-structures. We employ micro-scale X-ray tomography (also known as X-ray microscopy) to gain meaningful insights into the deformation and failure of these materials under compressive loads.

## 2. Materials and Methods

### 2.1. Ti$_2$AlC foam

In order to fabricate $Ti_2AlC$ foams in three distinct mesostructures, an appropriate precursor MAX phase powder was used in conjunction with sodium chloride pore formers following reported protocols [38, 41]. In this process $Ti_2AlC$ powder (Maxthal 211, Sandvik, Sweden) with a particle size in the range 45–90 μm, and three types of NaCl powders (Sigma-Aldrich, USA), with particle size distributions in the ranges 45–90 μm, 180–250 μm or 355–500 μm, were employed. The fabrication of foams was conducted in three main steps: (i) a mixture of the NaCl pore former (either coarse, medium or fine) and the $Ti_2AlC$ powder was blended in a 40/60 volume ratio by ball milling and then pressed in a cylindrical die of 12.7mm diameter at 800MPa; (ii) the NaCl pore former was dissolved in distilled water by soaking overnight, and (iii) the porous green $Ti_2AlC$ body was sintered under flowing argon at 1400°C for 4 hours. Pore sizes in the foams were determined by measuring the size of 50 pores in SEM images using the intercept method, as specified in ASTM E112-13 [42], from four SEM images in randomly selected locations on each sample.

### 2.2. Composite fabrication

To prepare composite Al/$Ti_2AlC$ specimens, Al alloy 6061 discs (McMaster-Carr, GA, USA) with a diameter of 20 mm and a thickness of 4 mm were used in an infiltration process. In this process the $Ti_2AlC$ foams with different pores sizes were "sandwiched" in between two Al alloy discs and placed in a graphite die, with graphite foils separating the discs from the die. This "sandwich" set-up facilitates a more uniform infiltration of molten metal. Infiltration was carried out using a spark plasma sintering system (SPS 25-10, GT Advanced Technologies, CA, USA). In this system, the chamber was evacuated and held at $10^{-6}$ torr for 10 minutes before heating. A direct current was pulsed at 10 ms intervals from 0 to 1250 A over 4 min to give a heating rate of 200 °C/minute before stabilizing at a current of 860 A to give a 1 min soak at 750 °C. The complete infiltration process including heating/melting, soaking, and cooling/solidification was carried out over 10 minutes. The temperature was calibrated and measured using procedures described elsewhere [27].

By employing pore formers with three different size distributions in the synthesis of $Ti_2AlC$ foams, we examined three distinct meso-structures of Al alloy/$Ti_2AlC$ composites, referred to as fine, medium and coarse, exhibiting Al alloy phase segments of different sizes. The term meso-structure is used as it exists at a level between the microstructure of the individual material constituents and the bulk composite macrostructure. Fine, medium and coarse composite samples were denoted by suffixes F-,M- and C- respectively, in this paper.

### 2.3. Characterization

Density and porosity (both open and closed) of all samples were determined by an alcohol immersion method, as specified in ASTM C20-00 [43]. Theoretical density values of 4.11 g/cm$^3$ and 2.70 g/cm$^3$ [44, 45] for $Ti_2AlC$ and Al alloy, respectively, were used to calculate a theoretical density of 3.55 g/cm$^3$ for composites containing 40 vol% Al, using the rule of mixture. Actual material densities are shown in Table 1.

To assess the influence of structure on the mechanical performance and failure of the composite materials, the deformation behaviour of materials in compression was tested. The compressive stress-strain curves of the specimens under cyclic loading were obtained using an MTS 810 (MTS Systems, MN, USA) servo hydraulic test frame at a strain rate of $7 \times 10^{-4}$ s$^{-1}$. All specimens for compressive testing were cut by electrical discharge machining to dimensions of 3.5 mm × 3.5 mm × 7 mm. All specimens were machined to have flat and parallel ends within ± 25 μm. As-processed and post-compression samples are denoted by the suffixes AP and PC, respectively.

Preliminary microstructural analysis of specimens was carried out using a Hitachi TM 3030 scanning electron microscope (SEM), with an accelerating voltage of 15 kV and a working distance of approximately 7 mm. The distribution of the different phases and the morphology of cracks in PC specimens were examined using this method. Electron backscatter diffraction (EBSD) analysis was further conducted to determine the phases present. EBSD scans were carried out using a Zeiss Ultra field emission gun SEM equipped with Oxford Instruments AZtec integrated EDS and EBSD system, with X-Max 20 mm2 silicon drift EDS detector and Nordlys-nano EBSD detector. The scans were done with a step size varying from 0.2 to 0.5 μm depending on the specimen (0.2 μm for the M-AP specimen, 0.3 μm for C-AP and 0.5 μm for F-AP). All the scans were done with an accelerating voltage of 20 keV. Data acquisition and analysis were done using Oxford Instrument's AZTec HKL and HKL Tango software.

To gain reliable insights into failure mechanisms in Al alloy/$Ti_2AlC$ composites, it is necessary to capture 3D microstructural information from as-processed and post-compression specimens. As cutting cross sections from specimens may result in further cracks, deformation and/or loss of material, it is preferable to conduct non-destructive 3D structural reconstruction. To achieve this, micro-scale X-ray computed tomography was carried out using a Zeiss MicroXCT-400. Micro-scale X-ray computed tomography, also known as $\mu$-CT, as carried out here constitutes a type of X-ray microscopy (XRM) as it uses optical elements in conjunction with a scintillator to acquire X-ray data at higher magnifications and spatial resolution [46, 47].

Several XRM scans with micrometre scale resolution were facilitated using a 150 kV, 10W X-ray beam. Specimens were rotated 360° in the sample chamber with 2D X-ray projections captured at 0.2° intervals with an exposure of 5 seconds per frame. Projections were captured using both a lens magnification of 4x and 20x and geometric positioning to result in pixels of linear size 4.95 and 1.01 $\mu$m, respectively. 2D X-ray projections were reconstructed using XMReconstructor software v7.0.2817 to yield isotropic voxels in 3D with the same linear dimensions. The reconstructed XRM data contain greyscale data representing material attenuation of X-rays. These data were interpreted using Avizo Fire 8 software (FEI Visualization Sciences Group). In this work we applied an entropic thresholding algorithm to quantitatively approximate the relative proportions of $Ti_2AlC$, Al alloy and air based on image greyscale histograms [48, 49]. An illustration of the thresholded phase distribution is shown in Figure 1. On the basis of 2D orthogonal slices and 3D reconstructions, the interaction of cracks with different phases was evaluated in post compression material. Volumetric segmentation analysis further facilitated the assessment of crack behaviour on the basis of the surface area to volume ratio.

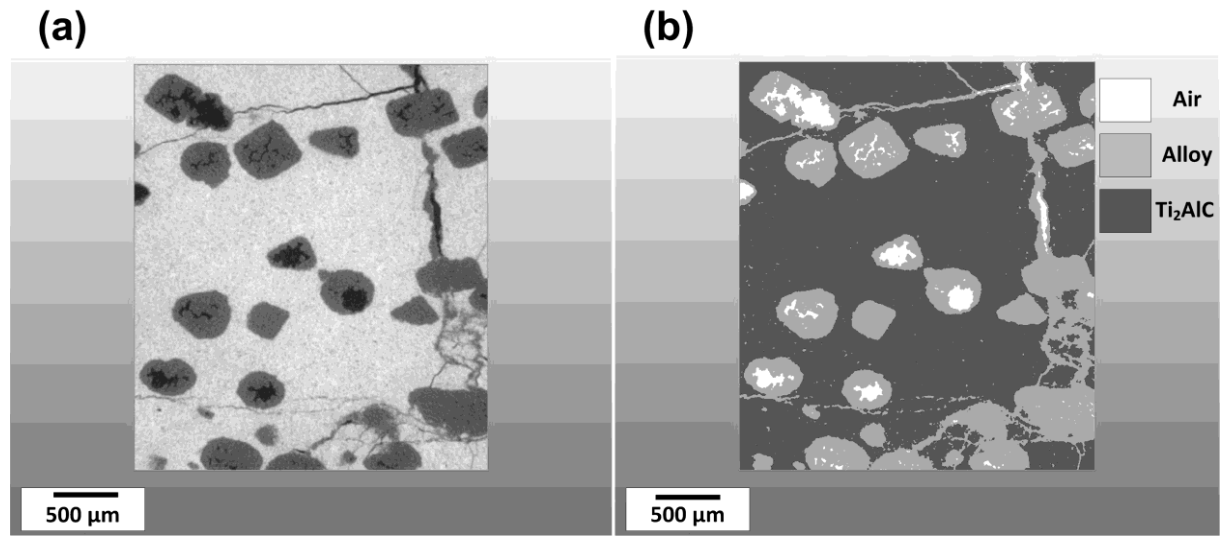

**Figure 1. Orthogonal slice from XRM based micro-CT (a) greyscale reconstruction results (b) three-phase thresholded data.**

### 3. Results
#### 3.1. Specimen properties

$Ti_2AlC$ foams with various pore sizes, *i.e.* 42–83 $\mu$m, 77–276 $\mu$m and 167–545 $\mu$m, were used for Al alloy infiltration to prepare Al alloy/$Ti_2AlC$ composites with fine, medium, and coarse structures, respectively. These structures were achieved using NaCl particles as pore formers and serve as precursors to fine, medium and coarse composite materials. All three foams exhibited connectivity of $Ti_2AlC$ grains and formation of sintering necks. These precursors were fabricated using the same volume percent (40 vol.%) of NaCl particles and thus have comparable overall porosities of 40.8, 41.6 and 39.9 vol.%. Thus the infiltration of these three foams with Al alloy resulted in roughly 40 vol% Al alloy/$Ti_2AlC$ composites with various interpenetrating phase sizes (Figures 2(a)–(c)).

**Table 1. Properties of composite materials with differing meso-structures.**

| Sample | Interpenetrating phase size[1], µm | Volume percent of constituents, vol.% | | | | Compressive strength[4], MPa | Failure strain[4], % |
|---|---|---|---|---|---|---|---|
| | | $Ti_2AlC$[2] vol% | Al alloy[3] vol% | Open pores[2] vol% | Closed pores[2] vol% | | |
| Fine | 42–83 | 60.1 ± 0.9 | 37.0 | 0.3 ± 0.2 | 2.6 ± 0.4 | 668 ± 28 | 1.27 ± 0.11 |
| Medium | 77–276 | 58.4 ± 1.2 | 35.9 | 0.5 ± 0.4 | 5.2 ± 0.3 | 610 ± 30 | 0.97 ± 0.06 |
| Coarse | 167–545 | 59.2 ± 1.4 | 34.6 | 0.7 ± 0.6 | 5.5 ± 0.4 | 563 ± 68 | 0.93 ± 0.04 |

[1]Determined from SEM images following ASTM E112-13 [42].
[2]Measured by alcohol immersion following ASTM C20-00 [43].
[3]Determined from the balance of $Ti_2AlC$ and pores.
[4]Averaged from 3 test specimens

Table 1 outlines the size of Al phase, volumetric contents of the constituents ($Ti_2AlC$, Al alloy), and open and closed porosity in fine, medium and coarse Al/$Ti_2AlC$ composites. Despite the short processing time, these data show that more than 97% of the open porosity in the foams was infiltrated with molten metal. The volume percent of closed pores in the composite materials is significantly higher than open pores. Closed pores most likely originate from shrinkage voids formed in the alloy during rapid cooling and pre-existing closed pores in the $Ti_2AlC$ foams prior to infiltration.

EBSD phase maps of the three different specimen types are given in Figure 2, illustrating the variation in mesostructure between the materials. From the presence of $Ti_3AlC_2$ and $Al_3Ti$ phases at interface regions it is evident that some reaction occurs between the MAX phase and the Al alloy during the infiltration process. These interface regions between the two phases are mainly composed of $Al_3Ti$ (shown in yellow) which most likely arises through Al diffusion from the liquid phase.

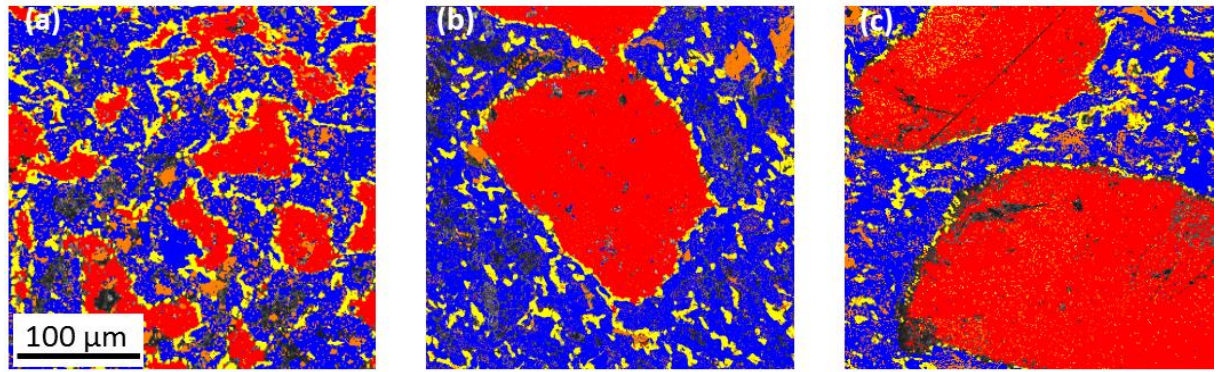

**Figure 2:** EBSD phase maps of (a) F-AP, (b) M-AP and (C) C-AP. Each colour represents a specific phase: red – Al; blue – $Ti_2AlC$; Orange – $Ti_3AlC_2$; yellow – $Al_3Ti$. All the maps are at the same magnification.

Stress-strain curves of the three Al/$Ti_2AlC$ composite materials tested to failure in compression are shown in Figure 3. Both strength and failure strain of the composites decrease with increasing interpenetrating phase size. The compressive strength and failure strain of the fine structured composites are approximately 20% and 35%, respectively, higher than those of the coarse/medium structured composites

.

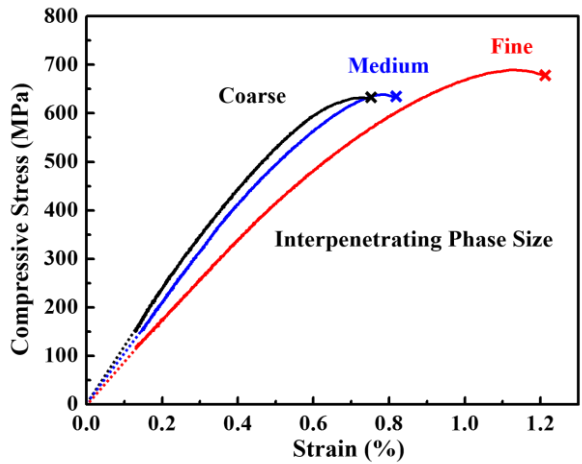

**Figure 3. Compressive stress/strain performance of Al alloy/Ti$_2$AlC composites with different meso-structures up to yield point**

## 3.2. SEM analysis

Figure 4(a) shows that the fine-structured material exhibited a single dominant crack subsequent to compressive yielding (F-PC), traversing throughout the entire material bulk. The diagonal crack plane exhibited the typical morphology of a shear type failure at roughly 45 ° relative to the loading direction, and exhibited step-like kinks over its length in which fracture debris appeared to be present, shown in Figure 4(b). This debris appeared to consist of a mixture of both Al alloy and Ti$_2$AlC phases as evident from the EDS results shown in Figure 5(a), with MAX phase grains exhibiting the layered microstructure typical of this material [12], shown in Figure 5(b).

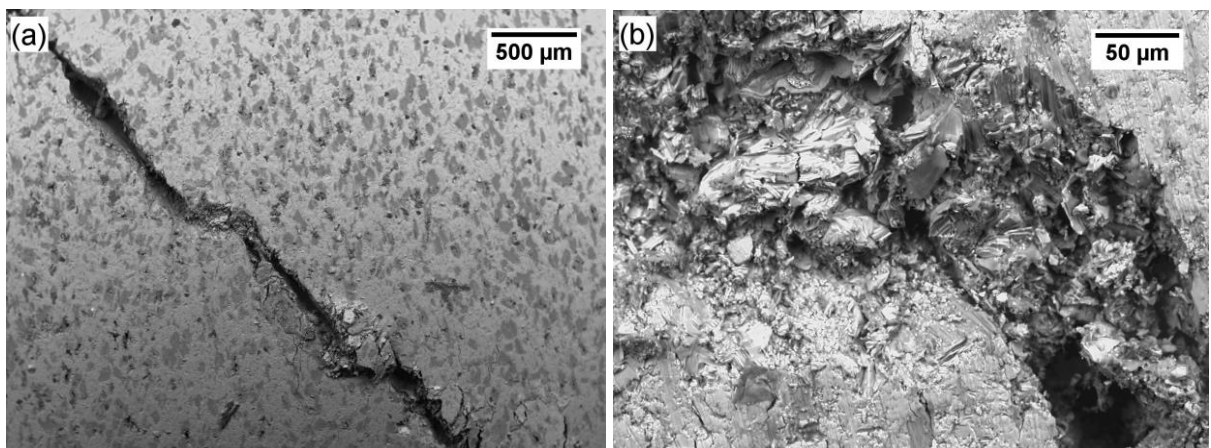

**Figure 4. SEM micrograph of F-PC material (a) low magnification (b) magnification of crack debris**

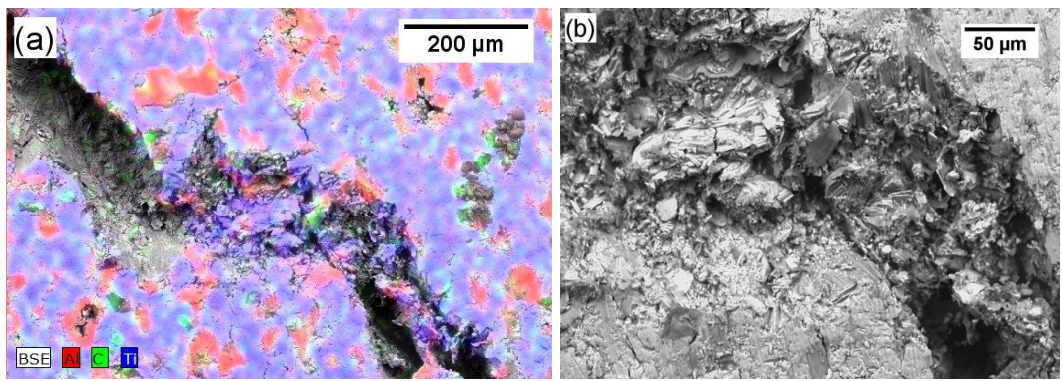

**Figure 5. SEM/EDS analysis of F-PC material (a) EDS spectrum of crack region (b) magnification of layered Ti$_2$AlC material**

In contrast to fine structured material, the medium and coarse structured materials did not exhibit a single dominant crack. Rather these materials exhibited a network of smaller cracks with some, as seen in Figure 6(a), exhibiting tearing propagating preferentially through the Al alloy/Ti$_2$AlC interfaces. The tendency of cracks to propagate around Al regions was observed in both coarse and medium structured material as well, shown in Figures 6(b) and 6(c), respectively.

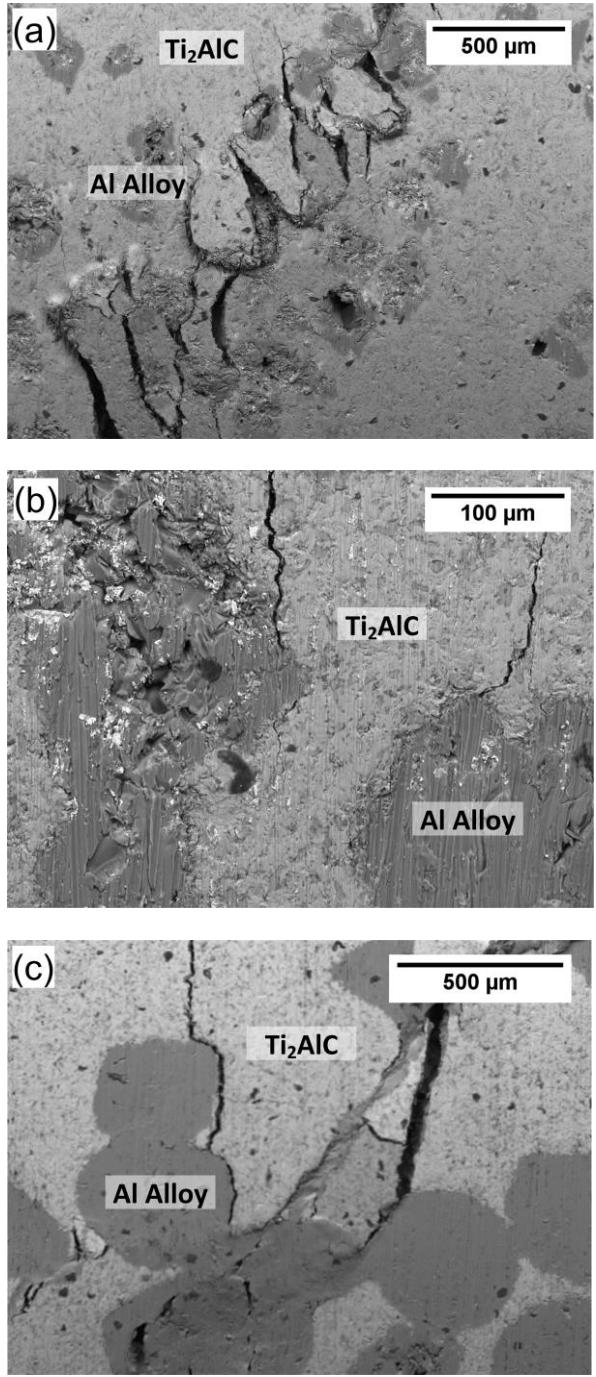

**Figure 6.** SEM micrographs of post compression materials (a) Jagged crack in M-PC (b) Crack interaction with Al in M-PC (c) Crack interaction with Al in C-PC

### 3.3. Tomography data

As SEM data only represents material surfaces, micro-scale X-ray tomography data is analysed in order to gain a meaningful insight into material structure and failure behaviour. Axial 2D slices were extracted from reconstructed X-ray data of all specimens. These ortho-slices, taken parallel and perpendicular to the loading direction, are consistent with propagation found from SEM analysis, showing crack propagation throughout the volume, rather than just at the surface, and shed further light on the structure and compressive yielding of the materials studied in the present work. The three phases that can be seen in the X-ray tomography data are, in order of increasing greyscale (and thus material attenuation), MAX phase, Al alloy and air. The latter, which appears as black or near black in the orthogonal slices shown in Figure 7, is apparent in cracks, shrinkage voids within Al regions and empty closed MAX phase pores that were not filled with Al. It was further found that cracks existed in the coarse precursor Ti$_2$AlC foams and were filled with Al during CAPAI. This resulted in elongated Al regions in as processed coarse structured (C-AP) materials as seen in Figure 7(d).

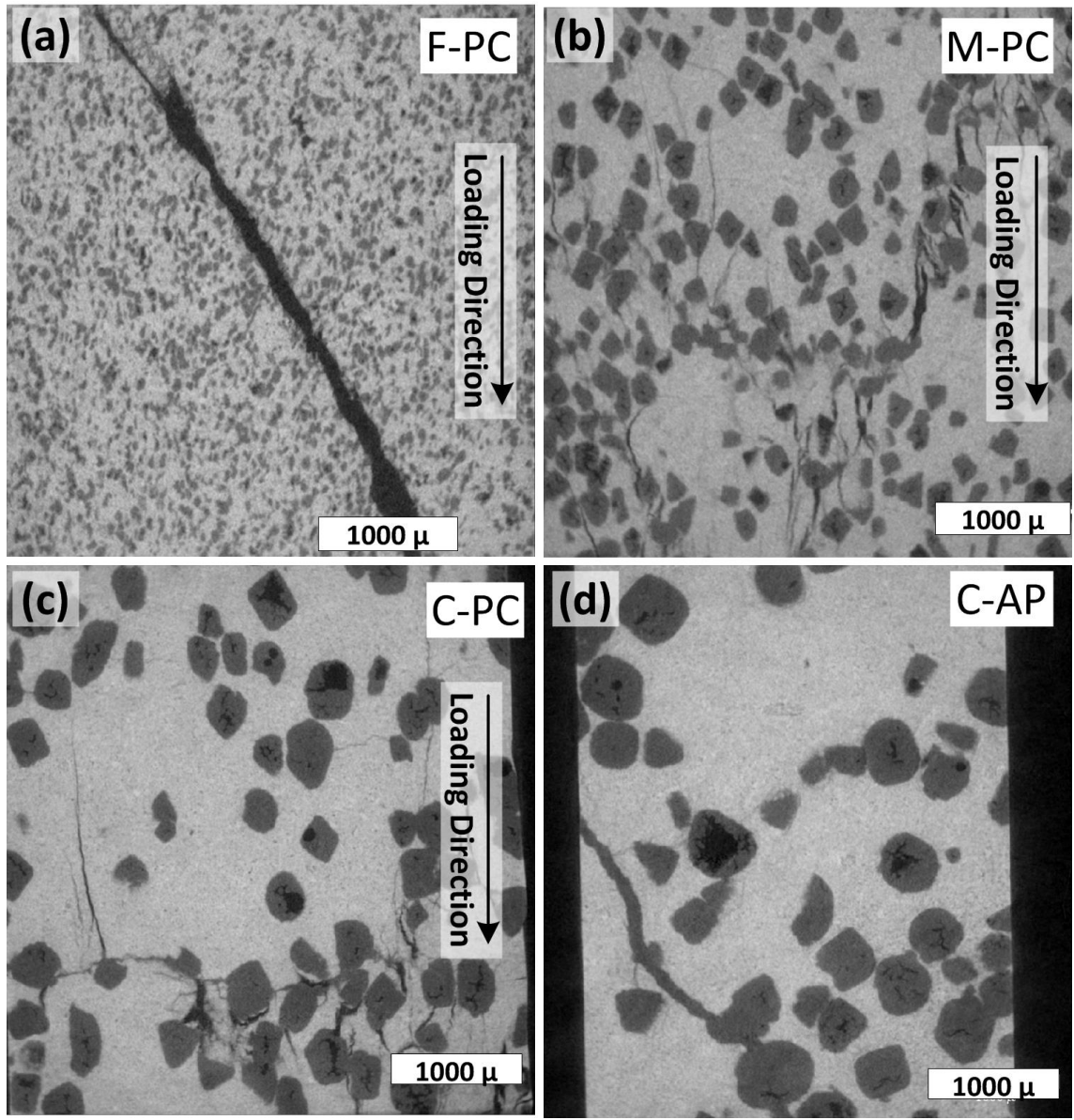

**Figure 7. XRM slices acquired at using a 4x objective lens, showing view perpendicular to loading direction (a) F-PC (b) M-CP (c) C-PC (d) C-AP**

On the basis of a review of parallel and transverse slices reviewed from each post-compression specimen, the interaction of propagating cracks with the MAX phase and Al alloy components was assessed. For medium and coarse structured specimens it was clearly evident that cracks propagate preferentially through the more brittle MAX phase rather than through the more ductile interpenetrating alloy, as illustrated by figures 6 and 7. Upon encountering Al filled pores, cracks generally tend to be deflected and propagate through MAX phase-Al alloy interfaces.

### 3.4. Volumetric X-ray analysis

Projections acquired from XRM were used to reconstruct volumetric representations of the porous MAX phase matrix constituent in fine-, medium- and coarse-structured Al/Ti$_2$ALC composites. These reconstructions shown in Figure 8 reveal the failed Ti$_2$AlC phase excluding the interpenetrating metal.

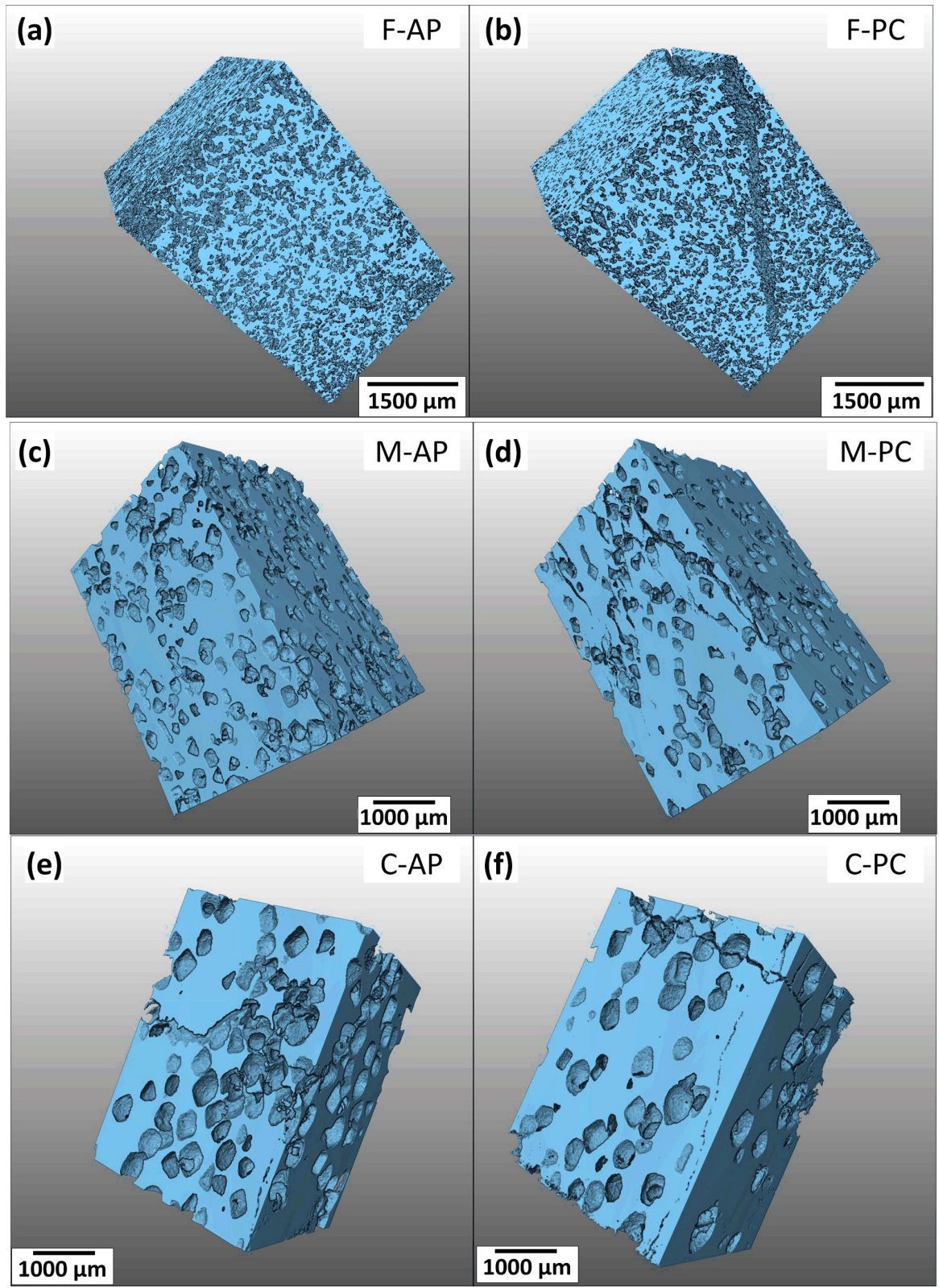

**Figure 8. Volume rendering of MAX phase**

On the basis of entropic thresholding, the approximated phase proportions of the three constituents (Al alloy, $Ti_2AlC$ and air) are shown in Table 2 for the different materials analysed. The post compression material shows a higher air fraction owing to the presence of cracks. The presence of unfilled pores and Al shrinkage voids further explains for the presence of air in both AP and PC materials. As entropic thresholding is an indirect method relative to the alcohol immersion method, the data given in Table 1 is likely to be a more accurate representation of as-processed materials. The largest void fraction is found for coarse material post-compression. This is likely the result of the wider cracks formed in this material during failure.

**Table 2. Phase proportions approximated by volumetric X-ray tomographic analysis**

| Material | Volumetric composition As-processed / Post Compression (%) | | |
|---|---|---|---|
| | $Ti_2AlC$ | Al 6061 | Air |
| Fine | 64.14/61.84 | 32.14/32.5 | 3.72/5.66 |
| Medium | 59.54/62.05 | 37.57/33.82 | 2.88/4.13 |
| Coarse | 59.56/63.81 | 38.05/30.25 | 2.38/5.94 |

### 3.5. Pore segmentation

The distribution of the aluminium alloy material was assessed by threshold-separating this phase and applying a watershed segmentation algorithm to define individual Al alloy-filled pores. The results of this segmentation are shown in Figure 9. The aluminium alloy phase is interconnected, as it is formed through an infiltration process in the predominantly open pore structure of the MAX phase foams, and thus segmentation is somewhat ambiguous. Nevertheless, the structure and the Al phase distribution are readily observable in the segmentation results. The volume and interface area of individual segments are naturally smaller for finer structures. The specific interfacial area of the alloy phase is represented by the parameter $α'$ given in terms of $μm^{-1}$ and corresponds to the ratio of interface area to pore volume.

$$\alpha' = \frac{\sum_{i=1}^{i=n} A_i}{\sum_{i=1}^{i=n} V_i} \ldots\ldots\ldots\ldots\ldots(1)$$

where $A_i$ and $V_i$ are the interfacial area and volume of individual segments. From the analysis of this parameter it is found unsurprisingly that the specific interface area of the Al alloy phase increases monotonically with finer mesostructures.

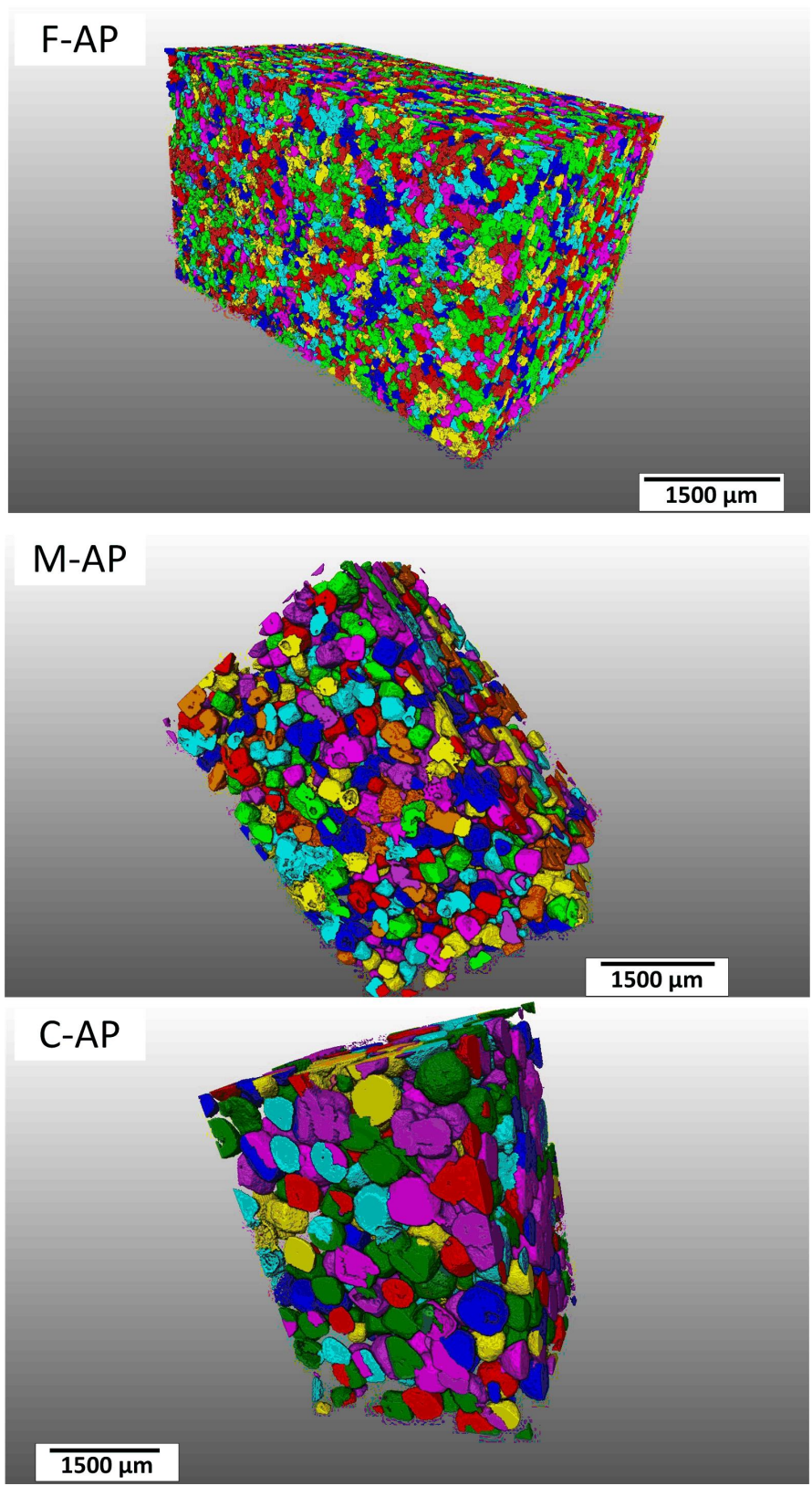

**Figure 9. Segmented volume rendering of Al alloy phase for fine, medium and coarse materials. Arbitrary segment colours.**

**Table 3. Aluminium phase segmentation**

| Material | Total interface ($\mu m^2$) | Mean segment volume ($\mu m^3$) | Mean segment interface ($\mu m^2$) | α' AP/PC ($\mu m^{-1}$) |
|---|---|---|---|---|
| **Fine** | $11.5 \times 10^8$ | $0.379 \times 10^6$ | $3.65 \times 10^4$ | $9.63 \times 10^{-2}$ / $10.78 \times 10^{-2}$ |
| **Medium** | $6.15 \times 10^8$ | $1.216 \times 10^6$ | $4.39 \times 10^4$ | $3.62 \times 10^{-2}$ / $4.48 \times 10^{-2}$ |
| **Coarse** | $3.92 \times 10^8$ | $3.241 \times 10^6$ | $8.10 \times 10^4$ | $2.50 \times 10^{-2}$ / $3.40 \times 10^{-2}$ |

## 4. Discussion

From examination of SEM micrographs, XRM slices, and volumetric analysis, it is evident that while the fine structure yields through a relatively straight shear crack in Al/Ti$_2$AlC processed by Al melt infiltration of the porous Ti$_2$AlC foams, the medium and coarse materials exhibit fine branching cracks. This is further evident from the quantitative analysis of phase segmentation. From analysis of the segmentation data we find that for fine, medium and coarse structured materials the specific interface area of the Al alloy phase increases subsequent to failure by approximately 12%, 24% and 36% respectively. This trend is indicative of two possible tendencies: (i) a greater extent of cracks; (ii) a preferential tendency for cracks to propagate through or around Al alloy inclusions. From the micrographic interpretation of ortho-slice sections there appears to be little tendency for preferential propagation within Al alloy regions. However an exception exists for the case of crack propagation through elongated Al alloy inclusions likely formed through the filling of pre-existing cracks in the MAX phase matrix, which are more common in coarse structured material. Thus it is likely that both mechanisms (i) and (ii) play a certain role in the observed trend.

While absent in the fine material, Al filled pores exhibiting large shrinkage voids are common in coarse and medium structures and these are likely to have acted as crack nucleation sites as evident by the correlation of lower compressive strength to higher porosity and the cracks seen in the XRM reconstruction of post compression materials. Such voids can be seen in XRM ortho-slices shown in Figure 7 (c) and (d). Minimising void formation in CAPAI synthesis is likely to improve toughness and may be achieved by employing a finer meso-structure or by increased infiltration pressure.

The tendency for cracks to propagate through Al alloy/Ti$_2$AlC interfaces suggests that fracture toughness can be improved by enhancing the bonding across these interfaces, thus increasing the tensile strength and energy dissipation. The unintentional formation of Al$_3$Ti and Ti$_3$AlC$_2$ near interfaces during infiltration, as revealed by EBSD, may weaken these regions and minimising their occurrence is likely to enhance fracture toughness. One method that may be appropriate towards this end would be the gas phase deposition of Al through PVD methods on MAX phase precursor to create a well-bonded intermediate layer with minimal diffusion impurities prior to CAPAI.

## 5. Conclusions

The compressive performance of a novel metal/MAX phase composite system was examined and interpreted with reference to crack propagation observations in materials of varied meso-structure. By employing recently developed techniques of X-ray microscopy we were able to gain three dimensional insights into the structure and failure behaviour of these materials. Understanding the role of meso-structure in the strength of metal/MAX phase composites provides new pathways towards the tailoring of mechanical properties in this type of system.

In the present work a finer interpenetrating metallic phase and lower porosity were found to strengthen the composite materials. This suggests that adjusting the size distribution of the metallic phase and minimising the void ratio in CAPAI processed materials may facilitate improved mechanical performance in these materials.

The major findings are summarized as follows.

1. While fine structured composite materials yields through a relatively straight shear crack, the medium and coarse materials exhibit yielding through large numbers of fine branching cracks.
2. Cracks propagate predominantly through the MAX phase. Upon encountering Al filled pores, cracks tend to be deflected and propagate through MAX phase-Al alloy interfaces.
3. The tendency for cracks to deflect through Al alloy/$Ti_2AlC$ interfaces suggests that fracture toughness can be improved by enhancing the bonding across these interfaces, thus increasing the tensile strength and energy dissipation that occurs in crack propagation.

**Acknowledgement**

We acknowledge access to XRM facilities of the Australian Microscopy & Microanalysis Research Facility at the Australian Centre for Microscopy & Microanalysis at the University of Sydney. This work was further supported by the U.S. Air Force Office of Scientific Research, MURI Program (FA9550-09-1-0686) and US National Science Foundation (NSF–1233792) to Texas A&M University. The authors would like to thank the program manager Dr. David Stargel for his support. In addition, the authors are also grateful for the support of the International Program Development Fund and DVC Research/International Research Collaboration Award, at the University of Sydney.